\documentstyle[12pt]{article}
\newcommand {\e} {\mbox{\rm e}}

\newcounter{eq}
\newcounter{sc}

\def\overleftrightarrow#1{\vbox{\ialign{##\crcr
 $\leftrightarrow$\crcr\noalign{\kern-1pt\nointerlineskip}
 $\hfil\displaystyle{#1}\hfil$\crcr}}}

\newlength{\minitwocolumn}
\begin{document}

%%%%%%%%%%%%%%%%%%%%%%%%%%%%%%%%%%%%%%%%%%%%%%%%%%%%%%%%%%%%%%%%%%
%%%%%%%%%%%%%%%%%%%%%%%% Title %%%%%%%%%%%%%%%%%%%%%%%%%%%%%%%%%%%
%%%%%%%%%%%%%%%%%%%%%%%%%%%%%%%%%%%%%%%%%%%%%%%%%%%%%%%%%%%%%%%%%%
\begin{flushright}
DPUR/TH/49\\
June, 2016\\
\end{flushright}
\vspace{20pt}

%\magnification=\magstep1
\pagestyle{empty}
\baselineskip15pt
%\font\cmssB=cmss17
%\font\cmssS=cmss10

\begin{center}
{\large\bf Fake Conformal Symmetry in Unimodular Gravity
\vskip 1mm }

\vspace{20mm}
Ichiro Oda \footnote{E-mail address:\ ioda@phys.u-ryukyu.ac.jp
}

\vspace{5mm}
           Department of Physics, Faculty of Science, University of the 
           Ryukyus,\\
           Nishihara, Okinawa 903-0213, Japan.\\

\end{center}

%\maketitle

\vspace{5mm}
\begin{abstract}
We study Weyl symmetry (local conformal symmetry) in unimodular gravity. 
It is shown that the Noether currents for both Weyl symmetry and global scale symmetry,
identically vanish as in the conformally invariant scalar-tensor gravity. We clearly explain
why in the class of conformally invariant gravitational theories, the Noether currents vanish 
by starting with the conformally invariant scalar-tensor gravity. 
Moreover, we comment on both classical and quantum-mechanical equivalences among Einstein's
general relativity, the conformally invariant scalar-tensor gravity and the Weyl-transverse
(WTDiff) gravity. Finally, we discuss the Weyl current in the conformally invariant scalar
action and see that it is also vanishing.
\end{abstract}

\newpage
\pagestyle{plain}
\pagenumbering{arabic}
%\setcounter{page}{1}

%%%%%%%%%%%%%%%%%%%%%%%%%%%%%%%%%%%%%%%%%%%%%%%%%%%%%%%%%%%%%%%%%%
%%%%%%%%%%%%%%%%%%%%%%%% Article %%%%%%%%%%%%%%%%%%%%%%%%%%%%%%%%%
%%%%%%%%%%%%%%%%%%%%%%%%%%%%%%%%%%%%%%%%%%%%%%%%%%%%%%%%%%%%%%%%%%

\rm
%%%%%%%%%%%%%%%%%%%%%%%%%%%%%%%%%%%%%%%%%%%%%%%%%%%%%%%%%%%%%%%%%%%%%
%%%%%%%%%%%%%%%%%%%%%%%%%%%%%%   SEC  1    %%%%%%%%%%%%%%%%%%%%%%%%%%
%%%%%%%%%%%%%%%%%%%%%%%%%%%%%%%%%%%%%%%%%%%%%%%%%%%%%%%%%%%%%%%%%%%%%
\section{Introduction}

One of the biggest mysteries in modern theoretical physics is certainly the problem of
the cosmological constant \cite{Weinberg}. This problem has several facets to be understood 
in a proper way. The $\it{old}$ cosmological constant problem is to understand simply 
why the vacuum energy density is so small. The vacuum energy density coming from 
gravitational fluctuations up to the Planck mass scale is larger than that observed experimentally 
by some 120 orders of magnitude. On the other hand, the $\it{new}$ cosmological constant problem 
is to understand why the cosmological constant is not exactly zero and why its energy density 
is the same order of magnitude as the present matter density.

Among several aspects of the cosmological constant problem, in particular, what the present
author would like to understand is the issue of radiative instability of the cosmological constant:
the necessity of fine-tuning the value of the cosmological constant every time the higher-order
loop corrections are added in perturbation theory. To resolve this problem, unimodular gravity
\cite{Einstein}-\cite{Padilla} has been put forward where the vacuum energy and $\it{a \ forteriori}$ 
all potential energy are decoupled from gravity since in the unimodular condition $\sqrt{-g} = 1$, 
the potential energy cannot couple to gravity at the action level.  

However, world is not so simple since in quantum field theories the unimodular condition must
be properly implemented via the Lagrange multiplier field, and then radiative corrections 
modify the Lagrange multiplier field, which essentially corresponds to the cosmological constant 
in unimodular gravity, thereby rendering its initial value radiatively unstable.

Thus, if unimodular gravity could provide us with some solution to the issue of radiative instability
of the cosmological constant, there should be a more symmetry or a still-unknown dynamical mechanism
to suppress radiative corrections to the vacuum energy. Actually, there has already appeared such
a theory where Weyl symmetry, i.e., the local conformal symmetry, is added to the volume preserving
diffeomorphisms or equivalently the transverse diffeomorphisms (TDiff) of unimodular gravity 
\cite{Izawa}-\cite{Alvarez2}.
We will henceforth call such the theory the Weyl-transverse (WTDiff) gravity. Many of
reasons why WTDiff gravity is better than the transverse gravity, which is called TDiff gravity,
are mentioned in Ref. \cite{Alvarez1}
where for instance, it is expected that if Weyl symmetry could survive even at the quantum level,
this theory would be a finite one although the one-loop calculation leads to anomalies 
in the Ward-Takahashi identities.   

The purposes of this article are three-fold. First, we calculate the Noether current for Weyl
symmetry in WTDiff gravity and show that it becomes identically vanishing as in the conformally
invariant scalar-tensor gravity \cite{Jackiw1}. Second, we provide a simple proof that the Weyl current 
vanishes in a class of conformally invariant gravitational theories. Finally, we generalize
this proof to a conformally invariant scalar matter action. 

This paper is organised as follows: In Section 2, we give two calculations of the Noether current 
for Weyl symmetry as in the same line of arguments as in Ref. \cite{Jackiw1}. In Section 3, we will 
demonstrate that three kinds of gravitational theories, those are, Einstein's general relativity,
the conformally invariant scalar-tensor gravity and WTDiff gravity are all equivalent at least 
classically, and comment on their quantum equivalences as well. 
In Section 4, we will clarify the reason why the Noether current made in Section 2 vanishes identically 
in this class of conformally invariant gravitational theories. In Section 5, we consider 
the Weyl-invariant scalar matter coupling with gravity and see that the Weyl Noether current
also vanishes in this case. The final section is devoted to discussions.

%%%%%%%%%%%%%%%%%%%%%%%%%%%%%%%%%%%%%%%%%%%%%%%%%%%%%%%%%%%%%%%%%%%%%
%%%%%%%%%%%%%%%%%%%%%%%%%%%%%%   SEC  2    %%%%%%%%%%%%%%%%%%%%%%%%%%
%%%%%%%%%%%%%%%%%%%%%%%%%%%%%%%%%%%%%%%%%%%%%%%%%%%%%%%%%%%%%%%%%%%%%
\section{The Noether current for Weyl symmetry}

We will start with the action of the Weyl-transverse (WTDiff) gravity in unimodular gravity 
\cite{Alvarez1, Alvarez2}, which is given by \footnote{We follow 
notation and conventions by Misner et al.'s textbook \cite{MTW}, for instance, the flat Minkowski metric
$\eta_{\mu\nu} = diag(-, +, +, +)$, the Riemann curvature tensor $R^\mu \ _{\nu\alpha\beta} = 
\partial_\alpha \Gamma^\mu_{\nu\beta} - \partial_\beta \Gamma^\mu_{\nu\alpha} + \Gamma^\mu_{\sigma\alpha} 
\Gamma^\sigma_{\nu\beta} - \Gamma^\mu_{\sigma\beta} \Gamma^\sigma_{\nu\alpha}$, 
and the Ricci tensor $R_{\mu\nu} = R^\alpha \ _{\mu\alpha\nu}$.
The reduced Planck mass is defined as $M_p = \sqrt{\frac{c \hbar}{8 \pi G}} = 2.4 \times 10^{18} GeV$.
Through this article, we adopt the reduced Planck units where we set $c = \hbar = M_p = 1$.}
%**   WTDiff Action 1  %%%%%%%%%%%%%%%%%%%%%%%%%%%%%%%%%%%%%%%%%%%%%%%%%%%%%%%%%
\begin{eqnarray}
S &=& \int d^4 x \ {\cal L}              \nonumber\\
&=& \frac{1}{12} \int d^4 x |g|^{\frac{1}{4}} \left[ R + \frac{3}{32} \frac{1}{|g|^2}
g^{\mu\nu} \partial_\mu |g| \partial_\nu |g|  \right],
\label{WTDiff Action 1}
\end{eqnarray}
%%%%%%%%%%%%%%%%%%%%%%%%%%%%%%%%%%%%%%%%%%%%%%%%%%%%%%%%%%%%%%%%%%%
where we have confined ourselves to four space-time dimensions since the generalization to
a general space-time dimension is straightforward. Moreover, we have selected the coefficient
$\frac{1}{12}$ for later convenience. Finally, note that we have defined as $g = \det g_{\mu\nu}
< 0$. 

This action (\ref{WTDiff Action 1}) turns out to be invariant under not the full group of
diffeomorphisms (Diff) but only the transverse diffeomorphisms (TDiff). Moreover, it is worthwhile
to notice that in spite of the existence of an explicit mass scale (the reduced Planck units 
which we have  set $M_p = 1$), this action is also invariant under Weyl transformation.
Actually, under the Weyl transformation
%**   Weyl transf  %%%%%%%%%%%%%%%%%%%%%%%%%%%%%%%%%%%%%%%%%%%%%%%%%%%%%%%%%
\begin{eqnarray}
g_{\mu\nu} \rightarrow g^\prime_{\mu\nu} = \Omega^2(x) g_{\mu\nu},
\label{Weyl transf}
\end{eqnarray}
%%%%%%%%%%%%%%%%%%%%%%%%%%%%%%%%%%%%%%%%%%%%%%%%%%%%%%%%%%%%%%%%%%%
the Lagrangian density in (\ref{WTDiff Action 1}) is changed as
%**   L'  %%%%%%%%%%%%%%%%%%%%%%%%%%%%%%%%%%%%%%%%%%%%%%%%%%%%%%%%%
\begin{eqnarray}
{\cal L}^\prime =  {\cal L} - \frac{1}{2} \partial_\mu \left( |g|^{\frac{1}{4}} 
g^{\mu\nu} \frac{1}{\Omega} \partial_\nu \Omega \right).
\label{L'}
\end{eqnarray}
%%%%%%%%%%%%%%%%%%%%%%%%%%%%%%%%%%%%%%%%%%%%%%%%%%%%%%%%%%%%%%%%%%%

Now we will calculate the Noether current for Weyl symmetry by using the Noether procedure \cite{Noether}.
We will closely follow the line of arguments in Ref. \cite{Jackiw1}. 
The general variation of the Lagrangian density in the action (\ref{WTDiff Action 1}) reads
%**   Var-L  %%%%%%%%%%%%%%%%%%%%%%%%%%%%%%%%%%%%%%%%%%%%%%%%%%%%%%%%%
\begin{eqnarray}
\delta {\cal L} =  \frac{\partial {\cal L}}{\partial g_{\mu\nu}} \delta g_{\mu\nu}
+ \frac{\partial {\cal L}}{\partial ( \partial_\mu g_{\nu\rho} )} \delta (\partial_\mu g_{\nu\rho})
+ \frac{\partial {\cal L}}{\partial ( \partial_\mu \partial_\nu g_{\rho\sigma} )} 
\delta ( \partial_\mu \partial_\nu g_{\rho\sigma} ).
\label{Var-L}
\end{eqnarray}
%%%%%%%%%%%%%%%%%%%%%%%%%%%%%%%%%%%%%%%%%%%%%%%%%%%%%%%%%%%%%%%%%%%
In this expression, let us note that the Lagrangian density at hand includes second-order derivatives 
of $g_{\mu\nu}$ in the scalar curvature $R$. Setting $\Omega(x) = \e^{- \Lambda(x)}$, the infinitesimal variation 
$\delta {\cal L}$ under the Weyl transformation (\ref{Weyl transf}) is given by
%**   delta L1  %%%%%%%%%%%%%%%%%%%%%%%%%%%%%%%%%%%%%%%%%%%%%%%%%%%%%%%%%
\begin{eqnarray}
\delta {\cal L} =  \partial_\mu X_1^\mu,
\label{delta L1}
\end{eqnarray}
%%%%%%%%%%%%%%%%%%%%%%%%%%%%%%%%%%%%%%%%%%%%%%%%%%%%%%%%%%%%%%%%%%%
where $X_1^\mu$ is defined as 
%**    X1  %%%%%%%%%%%%%%%%%%%%%%%%%%%%%%%%%%%%%%%%%%%%%%%%%%%%%%%%%
\begin{eqnarray}
X_1^\mu = \frac{1}{2} |g|^{\frac{1}{4}} g^{\mu\nu} \partial_\nu \Lambda.
\label{X1}
\end{eqnarray}
%%%%%%%%%%%%%%%%%%%%%%%%%%%%%%%%%%%%%%%%%%%%%%%%%%%%%%%%%%%%%%%%%%%
The equation (\ref{delta L1}) of course means that the action (\ref{WTDiff Action 1}) is invariant
under the Weyl transformation up to a surface term.

Next, using the equations of motion
%**   Eq.M  %%%%%%%%%%%%%%%%%%%%%%%%%%%%%%%%%%%%%%%%%%%%%%%%%%%%%%%%%
\begin{eqnarray}
\frac{\partial {\cal L}}{\partial g_{\mu\nu}} 
= \partial_\rho \frac{\partial {\cal L}}{\partial ( \partial_\rho g_{\mu\nu} )}
- \partial_\rho \partial_\sigma \frac{\partial {\cal L}}{\partial ( \partial_\rho \partial_\sigma g_{\mu\nu} )}, 
\label{Eq.M}
\end{eqnarray}
%%%%%%%%%%%%%%%%%%%%%%%%%%%%%%%%%%%%%%%%%%%%%%%%%%%%%%%%%%%%%%%%%%%
the variation $\delta {\cal L}$ in (\ref{Var-L}) can be cast to the form
%**   delta L2  %%%%%%%%%%%%%%%%%%%%%%%%%%%%%%%%%%%%%%%%%%%%%%%%%%%%%%%%%
\begin{eqnarray}
\delta {\cal L} =  \partial_\mu K_1^\mu,
\label{delta L2}
\end{eqnarray}
%%%%%%%%%%%%%%%%%%%%%%%%%%%%%%%%%%%%%%%%%%%%%%%%%%%%%%%%%%%%%%%%%%%
where $K_1^\mu$ is defined as 
%**    K1  %%%%%%%%%%%%%%%%%%%%%%%%%%%%%%%%%%%%%%%%%%%%%%%%%%%%%%%%%
\begin{eqnarray}
K_1^\mu = \frac{\partial {\cal L}}{\partial ( \partial_\mu g_{\nu\rho} )} \delta g_{\nu\rho}
+ \frac{\partial {\cal L}}{\partial ( \partial_\mu \partial_\nu g_{\rho\sigma} )} \partial_\nu \delta g_{\rho\sigma}
- \partial_\nu \frac{\partial {\cal L}}{\partial ( \partial_\mu \partial_\nu g_{\rho\sigma} )} \delta g_{\rho\sigma}.
\label{K1}
\end{eqnarray}
%%%%%%%%%%%%%%%%%%%%%%%%%%%%%%%%%%%%%%%%%%%%%%%%%%%%%%%%%%%%%%%%%%%
Using this formula, an explicit calculation yields
%**    K1=X1  %%%%%%%%%%%%%%%%%%%%%%%%%%%%%%%%%%%%%%%%%%%%%%%%%%%%%%%%%
\begin{eqnarray}
K_1^\mu = X_1^\mu,
\label{K1=X1}
\end{eqnarray}
%%%%%%%%%%%%%%%%%%%%%%%%%%%%%%%%%%%%%%%%%%%%%%%%%%%%%%%%%%%%%%%%%%%
thereby giving us the result that the Noether current for Weyl symmetry vanishes identically 
%**    J1  %%%%%%%%%%%%%%%%%%%%%%%%%%%%%%%%%%%%%%%%%%%%%%%%%%%%%%%%%
\begin{eqnarray}
J_1^\mu = K_1^\mu - X_1^\mu = 0.
\label{J1}
\end{eqnarray}
%%%%%%%%%%%%%%%%%%%%%%%%%%%%%%%%%%%%%%%%%%%%%%%%%%%%%%%%%%%%%%%%%%%
Let us note that the both expressions $X_1^\mu$ and $K_1^\mu$ are gauge invariant under the Weyl transformation.
This fact will be utilized later when we discuss a proof in Section 4.

As an alternative derivation of the same result, one can also appeal to a more conventional
method where the Lagrangian density in (\ref{WTDiff Action 1}) does not explicitly
involve second-order derivatives of $g_{\mu\nu}$ in the curvature scalar $R$.
To do that, one makes use of the following well-known formula: when one writes the scalar curvature
%**    R=R1+R2  %%%%%%%%%%%%%%%%%%%%%%%%%%%%%%%%%%%%%%%%%%%%%%%%%%%%%%%%%
\begin{eqnarray}
R = R_1 +  R_2,
\label{R=R1+R2}
\end{eqnarray}
%%%%%%%%%%%%%%%%%%%%%%%%%%%%%%%%%%%%%%%%%%%%%%%%%%%%%%%%%%%%%%%%%%%
the formula takes the form \cite{Fujii}
%**    Ident1  %%%%%%%%%%%%%%%%%%%%%%%%%%%%%%%%%%%%%%%%%%%%%%%%%%%%%%%%%
\begin{eqnarray}
R_1 = -2 R_2 + \frac{1}{\sqrt{-g}} \partial_\mu (\sqrt{-g} A^\mu),
\label{Ident1}
\end{eqnarray}
%%%%%%%%%%%%%%%%%%%%%%%%%%%%%%%%%%%%%%%%%%%%%%%%%%%%%%%%%%%%%%%%%%%
where one has defined the following quantities
%**    R&A  %%%%%%%%%%%%%%%%%%%%%%%%%%%%%%%%%%%%%%%%%%%%%%%%%%%%%%%%%
\begin{eqnarray}
R_1 &=& g^{\mu\nu} \left( \partial_\rho \Gamma^\rho_{\mu\nu} 
- \partial_\nu \Gamma^\rho_{\mu\rho} \right), \nonumber\\
R_2 &=& g^{\mu\nu} \left( \Gamma^\sigma_{\rho\sigma} \Gamma^\rho_{\mu\nu} 
- \Gamma^\sigma_{\rho\nu} \Gamma^\rho_{\mu\sigma} \right) \nonumber\\
&=& g^{\mu\nu} \Gamma^\sigma_{\rho\sigma} \Gamma^\rho_{\mu\nu} 
+ \frac{1}{2} \Gamma^\rho_{\mu\nu} \partial_\rho g^{\mu\nu}, \nonumber\\
A^\mu &=& g^{\nu\rho} \Gamma^\mu_{\nu\rho} - g^{\mu\nu} \Gamma^\rho_{\nu\rho}. 
\label{R&A}
\end{eqnarray}
%%%%%%%%%%%%%%%%%%%%%%%%%%%%%%%%%%%%%%%%%%%%%%%%%%%%%%%%%%%%%%%%%%%
Here let us note that $R_2$ is free of second-order derivatives of $g_{\mu\nu}$,
which are now involved in the term including $A^\mu$.

Then, we have the Lagrangian density
%**   L0-1  %%%%%%%%%%%%%%%%%%%%%%%%%%%%%%%%%%%%%%%%%%%%%%%%%%%%%%%%%
\begin{eqnarray}
{\cal L} = {\cal L}_0 + \frac{1}{12} \partial_\mu \left( |g|^{\frac{1}{4}} A^\mu \right),
\label{L0-1}
\end{eqnarray}
%%%%%%%%%%%%%%%%%%%%%%%%%%%%%%%%%%%%%%%%%%%%%%%%%%%%%%%%%%%%%%%%%%%
where ${\cal L}_0$ is defined as
%**   L0-2  %%%%%%%%%%%%%%%%%%%%%%%%%%%%%%%%%%%%%%%%%%%%%%%%%%%%%%%%%
\begin{eqnarray}
{\cal L}_0 = \frac{1}{12} |g|^{\frac{1}{4}} \left[ - R_2 + \frac{1}{4} |g|^{-1} A^\mu \partial_\mu |g|
+ \frac{3}{32} |g|^{-2} g^{\mu\nu} \partial_\mu |g| \partial_\nu |g| \right].
\label{L0-2}
\end{eqnarray}
%%%%%%%%%%%%%%%%%%%%%%%%%%%%%%%%%%%%%%%%%%%%%%%%%%%%%%%%%%%%%%%%%%%

We are now ready to show that the Noether current for Weyl symmetry is also zero by the more
conventional method. First of all, let us observe that the variation of ${\cal L}$ under the 
Weyl transformation (\ref{Weyl transf}) comes from only the total derivative term
%**   delta L3  %%%%%%%%%%%%%%%%%%%%%%%%%%%%%%%%%%%%%%%%%%%%%%%%%%%%%%%%%
\begin{eqnarray}
\delta {\cal L} = \partial_\mu \left( \frac{1}{2} |g|^{\frac{1}{4}} g^{\mu\nu} \partial_\nu \Lambda \right)
= \frac{1}{12} \partial_\mu \left[ \delta( |g|^{\frac{1}{4}} A^\mu ) \right].
\label{delta L3}
\end{eqnarray}
%%%%%%%%%%%%%%%%%%%%%%%%%%%%%%%%%%%%%%%%%%%%%%%%%%%%%%%%%%%%%%%%%%%
The total derivative terms are irrelevant to dynamics so in what follows let us focus our attention only on
the Lagrangian ${\cal L}_0$, which is free of second-order derivatives of $g_{\mu\nu}$.

Second, by an explicit calculation we find that the Lagrangian ${\cal L}_0$ is invariant under the Weyl 
transformation without any surface terms
%**    X2  %%%%%%%%%%%%%%%%%%%%%%%%%%%%%%%%%%%%%%%%%%%%%%%%%%%%%%%%%
\begin{eqnarray}
X_2^\mu = 0.
\label{X2}
\end{eqnarray}
%%%%%%%%%%%%%%%%%%%%%%%%%%%%%%%%%%%%%%%%%%%%%%%%%%%%%%%%%%%%%%%%%%%

Finally, applying the Noether theorem \cite{Noether} for ${\cal L}_0$, we can derive the following result
%**    K2  %%%%%%%%%%%%%%%%%%%%%%%%%%%%%%%%%%%%%%%%%%%%%%%%%%%%%%%%%
\begin{eqnarray}
K_2^\mu = \frac{\partial {\cal L}_0}{\partial ( \partial_\mu g_{\nu\rho} )} ( -2 g_{\nu\rho} ) = 0.
\label{K2}
\end{eqnarray}
%%%%%%%%%%%%%%%%%%%%%%%%%%%%%%%%%%%%%%%%%%%%%%%%%%%%%%%%%%%%%%%%%%%
Hence, the Noether current for Weyl symmetry identically vanishes
%**    J2  %%%%%%%%%%%%%%%%%%%%%%%%%%%%%%%%%%%%%%%%%%%%%%%%%%%%%%%%%
\begin{eqnarray}
J_2^\mu = K_2^\mu - X_2^\mu = 0.
\label{J2}
\end{eqnarray}
%%%%%%%%%%%%%%%%%%%%%%%%%%%%%%%%%%%%%%%%%%%%%%%%%%%%%%%%%%%%%%%%%%%
This result is very similar to that of the conformally invariant scalar-tensor gravity \cite{Jackiw1}.
This fact suggests that there might be a more universal proof which is independent of the form of actions
but reflects only the conformal invariance in this class of the Weyl-invariant gravitational theories.
In Section 4, we will present such a proof.

%%%%%%%%%%%%%%%%%%%%%%%%%%%%%%%%%%%%%%%%%%%%%%%%%%%%%%%%%%%%%%%%%%%%%
%%%%%%%%%%%%%%%%%%%%%%%%%%%%%%   SEC  3    %%%%%%%%%%%%%%%%%%%%%%%%%%
%%%%%%%%%%%%%%%%%%%%%%%%%%%%%%%%%%%%%%%%%%%%%%%%%%%%%%%%%%%%%%%%%%%%%
\section{Classical equivalence}

In this section, we wish to show the classical equivalence among the three kinds of gravitational theories,
those are, Einstein's general relativity, the conformally invariant scalar-tensor gravity and
WTDiff gravity. This demonstration of the equivalence of the three theories is necessary to present in
the next section the reason why the Noether current for Weyl symmetry, which was constructed in 
the previous section, identically vanishes.  

Incidentally, in this article, we are interested in not TDiff gravity but WTDiff one, both of which belong to
unimodular gravity. In addition to the problem of radiative instability of the Lagrange multiplier field
as mentioned in Section 1, one reason why we are not interested in TDiff gravity is that in TDiff gravity,
$g = \det g_{\mu\nu}$ becomes a dimensionless scalar field under TDiff as can be seen in the transformation
law $\delta g = \varepsilon^\mu \partial_\mu g$ under TDiff. As a result, any polynomials of $g$ are not excluded
by symmetries and allowed in principle to exist in the action. This fact makes it difficult to construct
the action consisting of a finite number of terms.

To show the equivalence, let us start with the Einstein-Hilbert action of general relativity
in four space-time dimensions
%**   EH Action 1  %%%%%%%%%%%%%%%%%%%%%%%%%%%%%%%%%%%%%%%%%%%%%%%%%%%%%%%%%
\begin{eqnarray}
S = \frac{1}{12} \int d^4 x \sqrt{-g} R.
\label{EH Action 1}
\end{eqnarray}
%%%%%%%%%%%%%%%%%%%%%%%%%%%%%%%%%%%%%%%%%%%%%%%%%%%%%%%%%%%%%%%%%%%
The well-known trick to enlarge gauge symmetries from Diff to WDiff is to introduce
the spurion field $\varphi$ and then construct a Weyl-invariant metric $\hat g_{\mu\nu} 
= \varphi^2 g_{\mu\nu}$ since under the Weyl transformation the spurion field transforms
as
%**   Weyl transf 2  %%%%%%%%%%%%%%%%%%%%%%%%%%%%%%%%%%%%%%%%%%%%%%%%%%%%%%%%%
\begin{eqnarray}
\varphi \rightarrow \varphi^\prime = \Omega^{-1}(x) \varphi.
\label{Weyl transf 2}
\end{eqnarray}
%%%%%%%%%%%%%%%%%%%%%%%%%%%%%%%%%%%%%%%%%%%%%%%%%%%%%%%%%%%%%%%%%%%
Replacing $g_{\mu\nu}$ with $\hat g_{\mu\nu}$ in the Einstein-Hilbert action, one can 
obtain the action of the conformally invariant scalar-tensor gravity \cite{Dirac, Deser}
%**   ST Action 1  %%%%%%%%%%%%%%%%%%%%%%%%%%%%%%%%%%%%%%%%%%%%%%%%%%%%%%%%%
\begin{eqnarray}
\hat S &=& \frac{1}{12} \int d^4 x \sqrt{- \hat g} \hat R   \nonumber\\
&=& \int d^4 x \sqrt{-g} \left[ \frac{1}{12} \varphi^2 R + \frac{1}{2} g^{\mu\nu} 
\partial_\mu \varphi \partial_\nu \varphi \right].
\label{ST Action 1}
\end{eqnarray}
%%%%%%%%%%%%%%%%%%%%%%%%%%%%%%%%%%%%%%%%%%%%%%%%%%%%%%%%%%%%%%%%%%%
Conversely, beginning by $\hat S$, to eliminate the spurion field $\varphi$ one can take
a gauge $\varphi = 1$ for Weyl symmetry, by which $\hat S$ is reduced to the Einstein-Hilbert 
action $S$ which is invariant only under Diff. In this sense, Einstein's general relativity
is classically equivalent to the conformally invariant scalar-tensor gravity.

Next let us show the equivalence between the conformally invariant scalar-tensor gravity and
WTDiff gravity. In this case, we start with the action $\hat S$ of the conformally invariant 
scalar-tensor gravity, and then take a different gauge condition $\varphi^2 = |g|^{-\frac{1}{4}}$,
i.e.,  
%**   WTDiff gauge  %%%%%%%%%%%%%%%%%%%%%%%%%%%%%%%%%%%%%%%%%%%%%%%%%%%%%%%%%
\begin{eqnarray}
\hat g_{\mu\nu} = \varphi^2 g_{\mu\nu} = |g|^{-\frac{1}{4}} g_{\mu\nu}.
\label{WTDiff gauge}
\end{eqnarray}
%%%%%%%%%%%%%%%%%%%%%%%%%%%%%%%%%%%%%%%%%%%%%%%%%%%%%%%%%%%%%%%%%%%
It is worthwhile to stress that this gauge condition is a gauge condition for
not the Weyl transformation but the longitudinal diffeomorphism. In fact, under the Weyl transformation
$|g|^{-\frac{1}{4}}$ transforms in the same way as the square of the spurion field does
%**   Weyl transf 3  %%%%%%%%%%%%%%%%%%%%%%%%%%%%%%%%%%%%%%%%%%%%%%%%%%%%%%%%%
\begin{eqnarray}
|g|^{-\frac{1}{4}} \rightarrow |g^\prime|^{-\frac{1}{4}} = \Omega^{-2}(x) |g|^{-\frac{1}{4}}.
\label{Weyl transf 3}
\end{eqnarray}
%%%%%%%%%%%%%%%%%%%%%%%%%%%%%%%%%%%%%%%%%%%%%%%%%%%%%%%%%%%%%%%%%%% 
Thus, the gauge condition (\ref{WTDiff gauge}) does not break Weyl symmetry, but breaks Diff down to TDiff
since 
%**   det hat g  %%%%%%%%%%%%%%%%%%%%%%%%%%%%%%%%%%%%%%%%%%%%%%%%%%%%%%%%%
\begin{eqnarray}
\hat g = \det \hat g_{\mu\nu} = -1.
\label{det hat g}
\end{eqnarray}
%%%%%%%%%%%%%%%%%%%%%%%%%%%%%%%%%%%%%%%%%%%%%%%%%%%%%%%%%%%%%%%%%%%
Now, substituting the gauge condition (\ref{WTDiff gauge}) into the action (\ref{ST Action 1}) of the
conformally invariant scalar-tensor gravity, it turns out that one arrives at the action (\ref{WTDiff Action 1}) 
of WTDiff gravity. Consequently, via the gauge-fixing procedure, the conformally invariant scalar-tensor gravity 
becomes equivalent to WTDiff gravity at least at the classical level. To summarize, we have found that the three
gravitational theories are classically equivalent via the trick of the introduction of the Weyl-invariant metric 
and the gauge-fixing procedure.

An important problem to ask is that the three gravitational theories, Einstein's general relativity, 
the conformally invariant scalar-tensor gravity and WTDiff gravity, are equivalent even at the quantum level.
To put differently, are there some anomalies, in particular, conformal anomaly for Weyl symmetry and a gauge 
anomaly for the longitudinal diffeomorphism?  We believe that there are no such anomalies by the following 
reasoning even if we never have any precise proof on this issue.       

In a pioneering work, Englert et al. \cite{Englert} have investigated the local conformal invariance 
in the conformally invariant scalar-tensor gravity. Their result is that as long as the local conformal
symmetry is spontaneously broken, anomalies do not arise. In this context, the spontaneous symmetry breakdown
of Weyl symmetry means that the spurion field takes the nonvanishing vacuum expectation value, 
$\langle \varphi(x) \rangle \neq 0$. Afterwards, this quantum equivalence has been studied from various
viewpoints and the affirmative answer has been obtained in \cite{Shaposhnikov1}-\cite{Ghilencea}. 
This observation could be also applied for the quantum equivalence
between the conformally invariant scalar-tensor gravity and WTDiff gravity although in this case the spurion field 
must take the vacuum expectation value which is not a constant but a field-dependent value. Anyway, we need
more studies to prove the exact equivalence among the three gravitational theories in future.

%%%%%%%%%%%%%%%%%%%%%%%%%%%%%%%%%%%%%%%%%%%%%%%%%%%%%%%%%%%%%%%%%%%%%
%%%%%%%%%%%%%%%%%%%%%%%%%%%%%%   SEC  4    %%%%%%%%%%%%%%%%%%%%%%%%%%
%%%%%%%%%%%%%%%%%%%%%%%%%%%%%%%%%%%%%%%%%%%%%%%%%%%%%%%%%%%%%%%%%%%%%
\section{Why is Noether current vanishing?}

In this section, on the basis of the results obtained in the previous section, we shall provide a simple proof
that the Noether current for Weyl symmetry in both the conformally invariant scalar-tensor gravity and WTDiff
gravity vanishes. For simplicity, we will consider the action which includes only first-order derivatives of
the metric tensor $g_{\mu\nu}$.

As the starting action, we will take the action $\hat S$ of the conformally invariant scalar-tensor gravity.
As in the case of WTDiff gravity, the action (\ref{ST Action 1}) can be rewritten in the first-order derivative form
%**   ST Action 2  %%%%%%%%%%%%%%%%%%%%%%%%%%%%%%%%%%%%%%%%%%%%%%%%%%%%%%%%%
\begin{eqnarray}
\hat S = \int d^4 x \left[ \hat {\cal L}_0 + \frac{1}{12} \partial_\mu ( \sqrt{-g} \varphi^2 A^\mu ) \right],
\label{ST Action 2}
\end{eqnarray}
%%%%%%%%%%%%%%%%%%%%%%%%%%%%%%%%%%%%%%%%%%%%%%%%%%%%%%%%%%%%%%%%%%%
where $\hat {\cal L}_0$ is defined by 
%**   hat-L0  %%%%%%%%%%%%%%%%%%%%%%%%%%%%%%%%%%%%%%%%%%%%%%%%%%%%%%%%%
\begin{eqnarray}
\hat {\cal L}_0 = \sqrt{-g} \left[ - \frac{1}{12} \varphi^2 R_2 
- \frac{1}{12} A^\mu \partial_\mu (\varphi^2)
+ \frac{1}{2} g^{\mu\nu} \partial_\mu \varphi \partial_\nu \varphi \right].
\label{hat-L0}
\end{eqnarray}
%%%%%%%%%%%%%%%%%%%%%%%%%%%%%%%%%%%%%%%%%%%%%%%%%%%%%%%%%%%%%%%%%%%
The total derivative term in $\hat S$ plays no role in bulk dynamics, so we will consider $\hat {\cal L}_0$
from now on. 

As shown in Ref. \cite{Jackiw1}, $\hat {\cal L}_0$ in invariant under the Weyl transfomation without
a surface term, from which we have
%**    X_0  %%%%%%%%%%%%%%%%%%%%%%%%%%%%%%%%%%%%%%%%%%%%%%%%%%%%%%%%%
\begin{eqnarray}
X_0^\mu = 0.
\label{X_0}
\end{eqnarray}
%%%%%%%%%%%%%%%%%%%%%%%%%%%%%%%%%%%%%%%%%%%%%%%%%%%%%%%%%%%%%%%%%%%
Then, the Noether theorem \cite{Noether} gives us
%**    K_0  %%%%%%%%%%%%%%%%%%%%%%%%%%%%%%%%%%%%%%%%%%%%%%%%%%%%%%%%%
\begin{eqnarray}
K_0^\mu = \frac{\partial \hat {\cal L}_0}{\partial ( \partial_\mu \varphi )} \varphi
+ \frac{\partial \hat {\cal L}_0}{\partial ( \partial_\mu g_{\nu\rho} )} ( -2 g_{\nu\rho} ).
\label{K_0}
\end{eqnarray}
%%%%%%%%%%%%%%%%%%%%%%%%%%%%%%%%%%%%%%%%%%%%%%%%%%%%%%%%%%%%%%%%%%%

We would like to prove $K_0^\mu = 0$, thereby leading to the vanishing Noether current
$J_0^\mu = 0$ owing to Eq. (\ref{X_0}). In fact, $K_0^\mu = 0$ has been
already shown by an explicit calculation \cite{Jackiw1}, but in this paper we will give a
simpler proof without much calculations. The key observation for our proof is to recall that
three kinds of gravitational theories are related to each other by a Weyl-invariant metric
$\hat g_{\mu\nu} = \varphi^2 g_{\mu\nu}$, from which taking the differentiation, we can 
derive the equation
%**    Key  %%%%%%%%%%%%%%%%%%%%%%%%%%%%%%%%%%%%%%%%%%%%%%%%%%%%%%%%%
\begin{eqnarray}
\partial_\mu \hat g_{\nu\rho} = 2 \varphi \partial_\mu \varphi g_{\nu\rho}
+ \varphi^2 \partial_\mu g_{\nu\rho}.
\label{Key}
\end{eqnarray}
%%%%%%%%%%%%%%%%%%%%%%%%%%%%%%%%%%%%%%%%%%%%%%%%%%%%%%%%%%%%%%%%%%% 
Using this equation, one finds that
%**    Key 2 %%%%%%%%%%%%%%%%%%%%%%%%%%%%%%%%%%%%%%%%%%%%%%%%%%%%%%%%%
\begin{eqnarray}
\frac{\partial \hat {\cal L}_0}{\partial ( \partial_\mu \varphi )}
&=& \frac{\partial \hat {\cal L}_0}{\partial ( \partial_\mu \hat g_{\nu\rho} )} 2 \varphi g_{\nu\rho},
\nonumber\\
\frac{\partial \hat {\cal L}_0}{\partial ( \partial_\mu g_{\nu\rho} )}
&=& \frac{\partial \hat {\cal L}_0}{\partial ( \partial_\mu \hat g_{\nu\rho} )} \varphi^2.
\label{Key 2}
\end{eqnarray}
%%%%%%%%%%%%%%%%%%%%%%%%%%%%%%%%%%%%%%%%%%%%%%%%%%%%%%%%%%%%%%%%%%% 
From Eq. (\ref{Key 2}), the equation (\ref{K_0}) produces the expected result
%**    K_0-2 %%%%%%%%%%%%%%%%%%%%%%%%%%%%%%%%%%%%%%%%%%%%%%%%%%%%%%%%%
\begin{eqnarray}
K_0^\mu = 0.
\label{K_0-2}
\end{eqnarray}
%%%%%%%%%%%%%%%%%%%%%%%%%%%%%%%%%%%%%%%%%%%%%%%%%%%%%%%%%%%%%%%%%%%
As a result, the Noether current for Weyl symmetry is vanishing
%**    J_0  %%%%%%%%%%%%%%%%%%%%%%%%%%%%%%%%%%%%%%%%%%%%%%%%%%%%%%%%%
\begin{eqnarray}
J_0^\mu = K_0^\mu - X_0^\mu = 0.
\label{J_0}
\end{eqnarray}
%%%%%%%%%%%%%%%%%%%%%%%%%%%%%%%%%%%%%%%%%%%%%%%%%%%%%%%%%%%%%%%%%%%

Accordingly, we have succeeded in giving a simpler proof that the Noether current for Weyl symmetry
is vanishing in the conformally invariant scalar-tensor gravity. Since the current is gauge invariant, 
our proof can be directly applied to any locally conformally invariant gravitational theories  such 
as WTDiff gravity etc., which are obtained via the trick $\hat g_{\mu\nu} = \varphi^2 g_{\mu\nu}$ 
and the gauge-fixing procedure from the diffeomorphism-invariant gravitational theories.

%%%%%%%%%%%%%%%%%%%%%%%%%%%%%%%%%%%%%%%%%%%%%%%%%%%%%%%%%%%%%%%%%%%%%
%%%%%%%%%%%%%%%%%%%%%%%%%%%%%%   SEC  4    %%%%%%%%%%%%%%%%%%%%%%%%%%
%%%%%%%%%%%%%%%%%%%%%%%%%%%%%%%%%%%%%%%%%%%%%%%%%%%%%%%%%%%%%%%%%%%%%
\section{Weyl-invariant matter coupling}

Since there are plenty of matters around us, it is natural to take account of effects of matter fields
in the present formalism.  In this section, we will show that the introduction of conformal matters
does not modify the fact that the Weyl current vanishes. As an example, we will work with the WDiff
coupling of a real scalar field with gravity, but the generalization to general matter fields is
straightforward as long as the matter fields are invariant under the Weyl transformation. 

As before, let us first begin by the action of a scalar field $\phi$ with a potential $V(\phi)$
in a curved space-time
%**   Scalar Matter 1  %%%%%%%%%%%%%%%%%%%%%%%%%%%%%%%%%%%%%%%%%%%%%%%%%%%%%%%%%
\begin{eqnarray}
S_m = \int d^4 x |g|^{\frac{1}{2}} \left[ - g^{\mu\nu} \partial_\mu \phi \partial_\nu \phi
- V(\phi)  \right].
\label{Scalar Matter 1}
\end{eqnarray}
%%%%%%%%%%%%%%%%%%%%%%%%%%%%%%%%%%%%%%%%%%%%%%%%%%%%%%%%%%%%%%%%%%%
Note that this action is manifestly invariant under the full group of diffeomorphisms (Diff). 
Under the Weyl transformation, the scalar field transforms as
%**   Weyl transf 4  %%%%%%%%%%%%%%%%%%%%%%%%%%%%%%%%%%%%%%%%%%%%%%%%%%%%%%%%%
\begin{eqnarray}
\phi \rightarrow \phi^\prime = \Omega^{-1}(x) \phi.
\label{Weyl transf 4}
\end{eqnarray}
%%%%%%%%%%%%%%%%%%%%%%%%%%%%%%%%%%%%%%%%%%%%%%%%%%%%%%%%%%%%%%%%%%%

The trick to enlarge gauge symmetries from Diff to WDiff is now to make both a Weyl-invariant metric 
$\hat g_{\mu\nu} = \varphi^2 g_{\mu\nu}$ and a Weyl-invariant scalar field $\hat \phi  = \varphi^{-1} \phi$, 
and then replace the metric and the scalar field in the action (\ref{Scalar Matter 1}) by the corresponding 
Weyl-invariant objects. As a result, the WDiff matter action takes the form
%**   Scalar Matter 2  %%%%%%%%%%%%%%%%%%%%%%%%%%%%%%%%%%%%%%%%%%%%%%%%%%%%%%%%%
\begin{eqnarray}
\hat S_m &=& \int d^4 x \ \hat {\cal L}_m                           \nonumber\\
&=& \int d^4 x |\hat g|^{\frac{1}{2}} \left[ - \hat g^{\mu\nu} \partial_\mu \hat \phi \partial_\nu \hat \phi
- V(\hat \phi)  \right]                                             \nonumber\\
&=& \int d^4 x |g|^{\frac{1}{2}} \left[ - \varphi^2 g^{\mu\nu} \partial_\mu \left(\frac{\phi}{\varphi}\right) 
\partial_\nu \left(\frac{\phi}{\varphi}\right) - \varphi^4 V\left(\frac{\phi}{\varphi}\right)  \right].
\label{Scalar Matter 2}
\end{eqnarray}
%%%%%%%%%%%%%%%%%%%%%%%%%%%%%%%%%%%%%%%%%%%%%%%%%%%%%%%%%%%%%%%%%%%

In this section, we shall calculate the Noether current for Weyl symmetry by the two different methods.
One method is to calculate the current in the WDiff matter action without gauge-fixing Weyl symmetry
like the conformally invariant scalar-tensor gravity. The other method is to gauge-fix the longitudinal
diffeomorphism by the gauge condition, by which the WDiff matter action is reduced to the WTDiff matter
one, and then calculate the Noether current for Weyl symmetry like the WTDiff gravity. The Weyl current is 
a gauge-invariant quantity, so both the methods should provide the same result.

First, let us calculate the Noether current for Weyl symmetry on the basis of the WDiff matter action 
(\ref{Scalar Matter 2}). It is easy to see that the action (\ref{Scalar Matter 2}) is invariant under 
the Weyl transformation without a surface term, which implies 
%**    X_m  %%%%%%%%%%%%%%%%%%%%%%%%%%%%%%%%%%%%%%%%%%%%%%%%%%%%%%%%%
\begin{eqnarray}
X_m^\mu = 0.
\label{X_m}
\end{eqnarray}
%%%%%%%%%%%%%%%%%%%%%%%%%%%%%%%%%%%%%%%%%%%%%%%%%%%%%%%%%%%%%%%%%%% 
Again, the Noether theorem \cite{Noether} yields
%**    K_m  %%%%%%%%%%%%%%%%%%%%%%%%%%%%%%%%%%%%%%%%%%%%%%%%%%%%%%%%%
\begin{eqnarray}
K_m^\mu = \frac{\partial \hat {\cal L}_m}{\partial ( \partial_\mu \phi )} \phi
+ \frac{\partial \hat {\cal L}_m}{\partial ( \partial_\mu \varphi )} \varphi
+ \frac{\partial \hat {\cal L}_m}{\partial ( \partial_\mu g_{\nu\rho} )} ( -2 g_{\nu\rho} ).
\label{K_m}
\end{eqnarray}
%%%%%%%%%%%%%%%%%%%%%%%%%%%%%%%%%%%%%%%%%%%%%%%%%%%%%%%%%%%%%%%%%%%
Next, the Weyl-invariant combinations $\hat g_{\mu\nu} = \varphi^2 g_{\mu\nu}$ and 
$\hat \phi  = \varphi^{-1} \phi$ give us the relations 
%**    Key 3 %%%%%%%%%%%%%%%%%%%%%%%%%%%%%%%%%%%%%%%%%%%%%%%%%%%%%%%%%
\begin{eqnarray}
\frac{\partial \hat {\cal L}_m}{\partial ( \partial_\mu \phi )}
&=& \frac{\partial \hat {\cal L}_m}{\partial ( \partial_\mu \hat \phi )} \frac{1}{\varphi},
\nonumber\\
\frac{\partial \hat {\cal L}_m}{\partial ( \partial_\mu \varphi )}
&=& \frac{\partial \hat {\cal L}_m}{\partial ( \partial_\mu \hat g_{\nu\rho} )} 2 \varphi g_{\nu\rho}
- \frac{\partial \hat {\cal L}_m}{\partial ( \partial_\mu \hat \phi )} \frac{\phi}{\varphi^2},
\nonumber\\
\frac{\partial \hat {\cal L}_m}{\partial ( \partial_\mu g_{\nu\rho} )}
&=& \frac{\partial \hat {\cal L}_m}{\partial ( \partial_\mu \hat g_{\nu\rho} )} \varphi^2.
\label{Key 3}
\end{eqnarray}
%%%%%%%%%%%%%%%%%%%%%%%%%%%%%%%%%%%%%%%%%%%%%%%%%%%%%%%%%%%%%%%%%%% 
Using these relations (\ref{Key 3}), $K_m^\mu$ in (\ref{K_m}) becomes zero 
%**    K_m-2 %%%%%%%%%%%%%%%%%%%%%%%%%%%%%%%%%%%%%%%%%%%%%%%%%%%%%%%%%
\begin{eqnarray}
K_m^\mu = 0.
\label{K_m-2}
\end{eqnarray}
%%%%%%%%%%%%%%%%%%%%%%%%%%%%%%%%%%%%%%%%%%%%%%%%%%%%%%%%%%%%%%%%%%%
The Noether current for Weyl symmetry is therefore vanishing
%**    J_m  %%%%%%%%%%%%%%%%%%%%%%%%%%%%%%%%%%%%%%%%%%%%%%%%%%%%%%%%%
\begin{eqnarray}
J_m^\mu = K_m^\mu - X_m^\mu = 0.
\label{J_m}
\end{eqnarray}
%%%%%%%%%%%%%%%%%%%%%%%%%%%%%%%%%%%%%%%%%%%%%%%%%%%%%%%%%%%%%%%%%%%
This is a general result and even after fixing the longitudinal diffeomorphism 
this result should be valid since the Weyl current is gauge invariant under the
Weyl transformation. Indeed, this is so by calculating the Weyl current in WTDiff
scalar action shortly.

Now let us take the gauge condition (\ref{WTDiff gauge}) for the longitudinal diffeomorphism,
which does not break the local conformal symmetry. Inserting the gauge condition (\ref{WTDiff gauge})
to the WDiff scalar matter action (\ref{Scalar Matter 2}) leads to the WTDiff scalar matter action
%**   Scalar Matter 3  %%%%%%%%%%%%%%%%%%%%%%%%%%%%%%%%%%%%%%%%%%%%%%%%%%%%%%%%%
\begin{eqnarray}
\hat S_m &=& \int d^4 x \ \hat {\cal L}_m                           \nonumber\\
&=& \int d^4 x \left[ - |g|^{\frac{1}{2}} g^{\mu\nu} \left( \frac{1}{64} \frac{\phi^2}{|g|^2} 
\partial_\mu |g| \partial_\nu |g| + \frac{1}{4} \frac{\phi}{|g|} \partial_\mu |g| \partial_\nu \phi
+ \partial_\mu \phi \partial_\nu \phi \right)  - V(|g|^{\frac{1}{8}} \phi)  \right].
\label{Scalar Matter 3}
\end{eqnarray}
%%%%%%%%%%%%%%%%%%%%%%%%%%%%%%%%%%%%%%%%%%%%%%%%%%%%%%%%%%%%%%%%%%%
  
Since the action (\ref{Scalar Matter 3}) is invariant under the Weyl transformation without a surface term, 
we have 
%**    X_m 2  %%%%%%%%%%%%%%%%%%%%%%%%%%%%%%%%%%%%%%%%%%%%%%%%%%%%%%%%%
\begin{eqnarray}
X_m^\mu = 0.
\label{X_m 2}
\end{eqnarray}
%%%%%%%%%%%%%%%%%%%%%%%%%%%%%%%%%%%%%%%%%%%%%%%%%%%%%%%%%%%%%%%%%%% 
The Noether theorem \cite{Noether} gives us the formula
%**    K_m 2 %%%%%%%%%%%%%%%%%%%%%%%%%%%%%%%%%%%%%%%%%%%%%%%%%%%%%%%%%
\begin{eqnarray}
K_m^\mu = \frac{\partial \hat {\cal L}_m}{\partial ( \partial_\mu \phi )} \phi
+ \frac{\partial \hat {\cal L}_m}{\partial ( \partial_\mu g_{\nu\rho} )} ( -2 g_{\nu\rho} ).
\label{K_m 2}
\end{eqnarray}
%%%%%%%%%%%%%%%%%%%%%%%%%%%%%%%%%%%%%%%%%%%%%%%%%%%%%%%%%%%%%%%%%%%
It is useful to evaluate each term in (\ref{K_m 2}) separately to see its gauge invariance
whose result is given by 
%**    K_m 3 %%%%%%%%%%%%%%%%%%%%%%%%%%%%%%%%%%%%%%%%%%%%%%%%%%%%%%%%%
\begin{eqnarray}
\frac{\partial \hat {\cal L}_m}{\partial ( \partial_\mu \phi )} \phi
&=& - \hat \phi^2 \hat g^{\mu\nu} \partial_\nu \log \left( \hat \phi^2 \right),   \nonumber\\
\frac{\partial \hat {\cal L}_m}{\partial ( \partial_\mu g_{\nu\rho} )} ( -2 g_{\nu\rho} )
&=& \hat \phi^2 \hat g^{\mu\nu} \partial_\nu \log \left( \hat \phi^2 \right).
\label{K_m 3}
\end{eqnarray}
%%%%%%%%%%%%%%%%%%%%%%%%%%%%%%%%%%%%%%%%%%%%%%%%%%%%%%%%%%%%%%%%%%%
As promised, each term is manifestly gauge invariant under the Weyl transformation since
it is expressed in terms of only gauge-invariant quantities. 
Adding the two terms in (\ref{K_m 3}), we have
%**    K_m 4 %%%%%%%%%%%%%%%%%%%%%%%%%%%%%%%%%%%%%%%%%%%%%%%%%%%%%%%%%
\begin{eqnarray}
K_m^\mu = 0.
\label{K_m 4}
\end{eqnarray}
%%%%%%%%%%%%%%%%%%%%%%%%%%%%%%%%%%%%%%%%%%%%%%%%%%%%%%%%%%%%%%%%%%%
Thus, the Noether current for Weyl symmetry is certainly vanishing
%**    J_m 2 %%%%%%%%%%%%%%%%%%%%%%%%%%%%%%%%%%%%%%%%%%%%%%%%%%%%%%%%%
\begin{eqnarray}
J_m^\mu = K_m^\mu - X_m^\mu = 0.
\label{J_m 2}
\end{eqnarray}
%%%%%%%%%%%%%%%%%%%%%%%%%%%%%%%%%%%%%%%%%%%%%%%%%%%%%%%%%%%%%%%%%%%
The both results in (\ref{J_m}) and (\ref{J_m 2}) clearly accounts for that the Noether current
for Weyl symmetry is vanishing in the conformally invariant scalar matter action as well.

%%%%%%%%%%%%%%%%%%%%%%%%%%%%%%%%%%%%%%%%%%%%%%%%%%%%%%%%%%%%%%%%%%%%%
%%%%%%%%%%%%%%%%%%%%%%%%%%%%%%   SEC  4    %%%%%%%%%%%%%%%%%%%%%%%%%%
%%%%%%%%%%%%%%%%%%%%%%%%%%%%%%%%%%%%%%%%%%%%%%%%%%%%%%%%%%%%%%%%%%%%%
\section{Discussions}

In this article, we have explicitly shown that the Noether currents in the Weyl-transverse 
(WTDiff) gravity, which is invariant under both the local conformal transformation and the transverse
diffeomorphisms, are identically vanishing for both local and global conformal symmetries.  Moreover, 
we have provided a simpler proof of the vanishing Weyl currents and stressed that all the locally 
conformally invariant gravitational theories, which are obtained via the trick $\hat g_{\mu\nu} 
= \varphi^2 g_{\mu\nu}$ from the diffeomorphism-invariant gravitational theories, have the vanishing 
Weyl currents.  We have also extended this calculation to the conformally invariant scalar matter theory
and shown that the Noether current is vanishing as well. 

This result of the vanishing Weyl currents in the conformally invariant scalar-tensor gravity and WTDiff
gravity is mathematically plausible since these two theories are at least classically equavalent to
Einstein's general relativity. In general relativity, there is no conformal invariance, so the Weyl
current is trivially zero and it is therefore also vanishing in the conformally invariant scalar-tensor gravity 
and WTDiff gravity.  In this sense, the local conformal symmetry existing in the conformally invariant 
scalar-tensor gravity and WTDiff gravity could be called a $\it{fake}$ conformal symmetry \cite{Jackiw1}. 
 
The important issue associated with the fake conformal symmetry is what advantage we have by
introducing a spurion field and adding the fake conformal symmetry to a quantum field theory. 
One opinion to this issue is that the fake Weyl invariance has no dynamical role and at
best a possible calculational device might be achieved \cite{Jackiw1}. \footnote{Similar or related 
criticisms are seen in \cite{Hertzberg, Quiros}.}
However, at least, there is one advantage in the sense that the conformally invariant scalar-tensor gravity
can be regarded as the more fundamental theory, from which via the gauge-fixing procedure, both general relativity
and WTDiff gravity can be derived in a natural way.  

Another opinion to this issue could be more interesting and hopeful. Since this fake conformal symmetry
is spurious in itself, this symmetry might not be broken by radiative corrections as advocated in 
various papers \cite{Englert}-\cite{Ghilencea}. \footnote{Related models with the fake conformal symmetry
are considered in Refs. \cite{Oda1}-\cite{Oda7}.} If this statement were true, the problem of the
radiative instability of the cosmological constant might be solved within the framework of theories
with the fake conformal symmetry. Of course, we need more investigations to check this interesting conjecture
in future.

%%%%%%%%%%%%%%%%%%%%%%%%%%%%%%%%%%%%%%%%%%%%%%%%%%%%%%%%%%%%%%%%%%
%%%%%%%%%%%%%%%%%%%%%%%% Acknowledgements %%%%%%%%%%%%%%%%%%%%%%%%%%%%%
%%%%%%%%%%%%%%%%%%%%%%%%%%%%%%%%%%%%%%%%%%%%%%%%%%%%%%%%%%%%%%%%%%
\begin{flushleft}
{\bf Acknowledgements}
\end{flushleft}
We wish to thank R. Percacci for discussions on quantum scale invariance
of the conformally-invariant scalar-tensor gravity.
This work is supported in part by the Grant-in-Aid for Scientific 
Research (C) No. 16K05327 from the Japan Ministry of Education, Culture, 
Sports, Science and Technology.

%%%%%%%%%%%%%%%%%%%%%%% reference %%%%%%%%%%%%%%%%%%%%%%%%%%%%%%%
%%%%%%%%%%%%%%%%%%%%%%%%%%%%%%%%%%%%%%%%%%%%%%%%%%%%%%%%%%%%%%%%%%

\end{document}